\documentclass[sn-aps,Numbered]{sn-jnl}

\usepackage[utf8]{inputenc}
\usepackage{amsmath}
\usepackage{amssymb}
\usepackage{bm}
\usepackage{enumerate}
\usepackage{graphicx}
\usepackage{hyperref}
\usepackage{lipsum}
\usepackage{slashed}
\usepackage{tensor}
\usepackage[dvipsnames]{xcolor}

\begin{document}

\title{Gravitational Collapse in Higher-Dimensional Rastall Gravity with and without Cosmological Constant}





\title{Gravitational Collapse in Higher-Dimensional Rastall Gravity with and without Cosmological Constant}

\author[1]{\fnm{Golfin} \sur{Ekatria}}\email{golfinekatria@gmail.com}

\author[2]{\fnm{Andy Octavian} \sur{Latief}}\email{latief@itb.ac.id}

\author[1]{\fnm{Fiki Taufik} \sur{Akbar}}\email{ftakbar@itb.ac.id}

\author*[1]{\fnm{Bobby Eka} \sur{Gunara}}\email{bobby@itb.ac.id}

\affil[1]{\orgdiv{Theoretical Physics Laboratory, Theoretical High Energy Physics Research Division, Faculty of Mathematics and Natural Sciences}, \orgname{Institut Teknologi Bandung}, \orgaddress{\street{Jl.~Ganesha no.~10}, \city{Bandung}, \postcode{40132}, \country{Indonesia}}}

\affil[2]{\orgdiv{Physics of Magnetism and Photonics Research Division, Faculty of Mathematics and Natural Sciences}, \orgname{Institut Teknologi Bandung}, \orgaddress{\street{Jl.~Ganesha no.~10}, \city{Bandung}, \postcode{40132}, \country{Indonesia}}}

\abstract{
    We consider a spherically symmetric homogeneous perfect fluid undergoing a gravitational collapse to singularity in the framework of higher-dimensional Rastall gravity in the cases of vanishing and nonvanishing cosmological constants. The possible final states of the collapse in any finite dimension are black hole and naked singularity, but the naked singularity formation becomes less favored when the dimension is increased. We find that there are two physically distinct solutions for the collapse evolution in the case of nonzero cosmological constant: trigonometric and exponential solutions. The effective energy density of the fluid is decreasing (increasing) in the former (latter) when the magnitude of the cosmological constant is increased, which implies that the former undergoes a slower collapse than the latter. Furthermore, we find that a temporary trapped surface is possible to emerge in the case of trigonometric solution in the naked singularity region only. Therefore, distant observers with observational time shorter than the collapse duration may conclude that a black hole is formed, although the collapse will eventually lead to a naked singularity formation.}

\maketitle

\section{Introduction}

\label{sec-introduction}

The final fate of massive stars undergoing gravitational collapse after exhausting completely their internal nuclear fuels has attracted significant attention in the last few decades. Unlike smaller stars that may reach equilibrium states in the form of neutron stars or white dwarf stars, the massive ones are predicted by Penrose-Hawking singularity theorems to collapse to smaller and smaller radius until spacetime singularities are formed \cite{Hawking-Ellis-large-2011}. The gravitational collapse of a spherical cloud of homogeneous dust was shown by Oppenheimer and Snyder \cite{Oppenheimer-Snyder-continued-1939}, and independently by Datt \cite{Datt-on-a-class-1938}, to lead to the formation of a black hole. This early model of gravitational collapse obeys the cosmic censorship conjecture (CCC) introduced by Penrose, which states that the final fate of gravitational collapse is always in the form of black hole \cite{Penrose-gravitational-2002}.

However, later developments have shown that the final state of gravitational collapse is either black hole or naked singularity, depending on whether the singularity is hidden behind the event horizon or not \cite{Joshi-Malafarina-recent-2011}, hence suggesting potential counterexamples to the CCC. The possibility of the formation of naked singularities has been demonstrated in various models, such as in the case of perfect fluids \cite{Harada-final-1998,da-Rocha-Wang-collapsing-2000,Harada-Maeda-convergence-2001,Giambo-et-al-naked-2003,Giambo-et-al-naked-2004}, imperfect fluids \cite{Lake-collapse-1982,Szekeres-Iyer-spherically-1993,Barve-et-al-spherical-2000}, and scalar fields \cite{Christodoulou-examples-1994,Brady-self-1995,Christodoulou-the-instability-1999,Giambo-gravitational-2005}. Other models that provide counterexamples to the CCC have been studied in the context of modified gravity theories, such as in $f(\mathcal{R})$ gravity \cite{Ziaie-et-al-naked-2011,Ghosh-Maharaj-gravitational-2012}, Gauss-Bonnet gravity \cite{Maeda-final-2006,Abbas-Tahir-gravitational-2017,Tavakoli-et-al-exploring-2021,Tavakoli-et-al-role-2022,Dialektopoulos-et-al-gravitational-2023}, Brans-Dicke theory \cite{Bedjaoui-et-al-existence-2010,Ziaie-et-al-naked-2010,Ziaie-et-al-gravitational-2024}, Eddington-inspired Born-Infeld theory \cite{Tavakoli-et-al-final-2017,Shaikh-Joshi-gravitational-2018}, and Lyra geometry \cite{Ziaie-et-al-trapped-2014}.

One of the important modified theories of gravity is the one introduced by P.~Rastall in 1972 \cite{Rastall-generalization-1972}, in which he proposed that the covariant divergence of the matter energy-momentum tensor is proportional to the covariant divergence of the Ricci scalar curvature, hence discarding the assumption of divergence-free energy-momentum tensor. As a generalization of Einstein’s general relativity theory, it is therefore important to investigate how the process of gravitational collapse might be different in the Rastall gravity. This problem has been studied by several works, for example Ref.~\cite{Ziaie-Tavakoli-null-2020} for the case of null fluid collapse and Ref.~\cite{Ziaie-et-al-gravitational-2019} for the case of perfect fluid collapse. In particular, in Ref.~\cite{Ziaie-et-al-gravitational-2019}, A.~Ziaie, H.~Moradpour, and S.~Ghaffari demonstrated that the final outcome of a spherically symmetric, homogeneous perfect fluid undergoing gravitational collapse can result in either a black hole or a naked singularity, depending on the values of the Rastall parameter and the barotropic index in the equation of state of the fluid. This finding contributes to the growing list of counterexamples to the CCC.

In this paper, we generalize the work in Ref.~\cite{Ziaie-et-al-gravitational-2019} to two directions. First is the generalization to higher dimensions, following several attempts in this direction \cite{Banerjee-et-al-gravitational-1994,Ilha-Lemos-dimensionally-1997,Ilha-et-al-dimensionally-1999,da-Rocha-Wang-collapsing-2000,Ghosh-Beesham-higher-2001,Ghosh-Dadhich-naked-2001,Ghosh-Banerjee-non-marginally-2003,Goswami-Joshi-spherical-2004,Goswami-Joshi-cosmic-2004,Debnath-et-al-quasi-spherical-2004,Debnath-Chakraborty-naked-2004,Mahajan-et-al-cosmic-2005,Goswami-Joshi-spherical-2007}. We map possible final states of gravitational collapse in the space of Rastall parameter and barotropic index, as in Ref.~\cite{Ziaie-et-al-gravitational-2019}, but now extending it to higher dimensions. We find that there are two distinct naked singularity regions in this parameter space: the one in the lower branch in Fig.~\ref{fig-parameter-space}, which contains the case of naked singularity formation in the general relativity limit, and the one in the upper branch. Previous studies within the framework of Einstein's general relativity have reported that the region corresponding to a naked singularity is monotonically shrinking when the spacetime dimension is increased, suggesting that the CCC may be restored in higher dimensions \cite{Ghosh-Dadhich-naked-2001,Goswami-Joshi-cosmic-2004,Goswami-Joshi-spherical-2004,Mahajan-et-al-cosmic-2005}. It is then natural to investigate whether the CCC is also restored in higher dimensions when we generalize the model to Rastall gravity by studying what happens to the naked singularity regions mentioned above. The extension to higher dimensions is further motivated by the emergence of novel phenomena arising from the incorporation of additional dimensions. For example, Refs.~\cite{Tavakoli-et-al-exploring-2021,Tavakoli-et-al-role-2022}, which examined gravitational collapse leading to a central singularity for a Dvali-Gabadadze-Porrati brane in five dimensions within the Gauss-Bonnet framework, reported the occurence of sudden singularity during the collapse, prior to the formation of the central singularity. This sudden singularity then can either be hidden by a black hole horizon or remain exposed, leading to the formation of a naked sudden singularity in the latter case. This intriguing phenomenon further motivates us to investigate the influence of higher dimensions on the evolution of gravitational collapse.

The second generalization, inspired by several works in this direction \cite{Garfinkle-Vuille-gravitational-1991,Cissoko-et-al-gravitational-1998,Lemos-collapsing-1999,Wagh-Maharaj-naked-1999,Markovic-Shapiro-gravitational-2000,Lake-gravitational-2000,Deshingkar-et-al-gravitational-2001,Goncalves-naked-2001,Ghosh-inhomogeneous-2005,Ghosh-Deshkar-higher-2007,Sharif-Ahmad-gravitational-2007,Abbas-et-al-expansion-2019}, is performed by incorporating the effects of the cosmological constant to the gravitational collapse. Here we focus only on the case where the final outcome of the gravitational collapse is a spacetime singularity, either in the form of black hole or naked singularity, unlike previous studies where the bounce takes place such that the singularity is not formed \cite{Markovic-Shapiro-gravitational-2000,Deshingkar-et-al-gravitational-2001,Madhav-et-al-gravitational-2005}. Although the cosmological constant is usually assumed to be of positive value in an attempt to model the dark energy in the universe \cite{Riess-et-al-observational-1998,Perlmutter-et-al-measurements-1999,Ashtekar-et-al-gravitational-2016,Ashtekar-implications-2017}, here we assume that it can be of negative value as well for the sake of theoretical completeness. The interior spacetime will be matched at the boundary of the star with the exterior spacetime endowed with the Vaidya-(anti-)de Sitter metric, obtained by generalizing the Vaidya metric \cite{Vaidya-Shah-1957} to higher dimensions \cite{Ghosh-Dadhich-naked-2001} and including the cosmological constant term \cite{Wagh-Maharaj-naked-1999}. With this setup, we then investigate the influence of the cosmological constant to the black hole and the naked singularity formations in the gravitational collapse.

We organize the paper as the following: In Sec.~\ref{sec-Rastall-field-equations} we derive the field equations of the Rastall gravity with nonvanishing cosmological constant for the interior spacetime of a spherically symmetric homogeneous perfect fluid star in order to find an equation that governs the time evolution of the star's gravitational collapse. We then discuss its solutions for the case of zero and nonzero cosmological constants in Secs.~\ref{sec-zero-Lambda} and \ref{sec-nonzero-Lambda}, respectively. We then conclude the paper in Sec.~\ref{sec-conclusions}.

\section{Rastall Gravity Field Equations}

\label{sec-Rastall-field-equations}

Let us first discuss the generalization of Rastall theory of gravity \cite{Rastall-generalization-1972} in higher dimensions. According to this theory, the local energy-momentum tensor $\mathcal{T}_{\mu \nu}$ is not conserved, and its covariant derivative is taken to be proportional to the covariant derivative of the Ricci curvature scalar,
\begin{equation}
    \nabla_\nu \mathcal{T}^{\mu \nu} = \lambda g^{\mu \nu} \nabla_\nu \mathcal{R}, \label{eq:RastallIdea}
\end{equation}
where $\lambda$ is the Rastall parameter, $g^{\mu \nu}$ is the inverse of the metric $g_{\mu \nu}$ endowed on a $d$-dimensional spacetime $\mathbb{M}^d$, and $\mathcal{R}$ is the Ricci scalar of $\mathbb{M}^d$. The gravitational Rastall field equation, including also the term proportional to the cosmological constant $\Lambda$, is given by
\begin{equation}
    \mathcal{G}^{\mu \nu} + \gamma g^{\mu \nu} \mathcal{R} + g^{\mu \nu} \Lambda = \kappa \mathcal{T}^{\mu \nu}, \label{eq:RastallField}
\end{equation}
where $\mathcal{G}_{\mu \nu}$ is the Einstein tensor and $\gamma = \kappa \lambda$ is Rastall dimensionless parameter, with
\begin{equation}
    \kappa = \frac{1}{2} \left( \frac{\eta}{\eta - 2 \gamma + 1} \right) \frac{8 \pi G}{c^4} \label{eq:kappa}
\end{equation}
and
\begin{equation}
    \eta \equiv (2 \gamma - 1) d + 2. \label{eq:eta}
\end{equation}
Eq.~\eqref{eq:RastallField} then can be written as
\begin{equation}
    \mathcal{G}_{\mu \nu} = \kappa \mathcal{S}_{\mu \nu}, \qquad \mathcal{S}_{\mu \nu} \equiv \mathcal{T}_{\mu \nu} - \frac{2 \gamma \mathcal{T} - (d - 2) \tilde{\Lambda}}{\eta} g_{\mu \nu}, \label{eq:RastallFieldEqForm}
\end{equation}
where $\mathcal{S}_{\mu\nu}$ is called the effective energy-momentum tensor and $\tilde{\Lambda} \equiv \Lambda/\kappa$. For the rest of the paper we restrict the discussion to the case of homogeneous and isotropic perfect fluid whose energy-momentum tensor has the form
\begin{equation}
    \mathcal{T}^{\mu \nu} = (\rho + p) U^\mu U^\nu + p g^{\mu \nu},
\end{equation}
where $\rho$ is the energy density, $p$ is the  pressure, and $U^2 \equiv U_\mu U^\mu = -1$. Thus, the components of the effective energy-momentum tensor are
\begin{eqnarray}
    \tensor{S}{^0_0} &\equiv& -\rho^\text{eff} = -\frac{1}{\eta} \Big\{ (\eta - 2 \gamma) \rho + \, 2 \gamma \left[ p_r + (d - 2) p_t \right] + (d - 2) \tilde{\Lambda} \Big\}, \label{eq:Rhoeff} \\
    \tensor{S}{^1_1} &\equiv& p_r^\text{eff} = \frac{1}{\eta} \Big\{ (\eta - 2 \gamma) p_r  + \, 2 \gamma \left[ \rho - (d - 2) p_t \right] - (d - 2) \tilde{\Lambda} \Big\}, \label{eq:Preff} \\
    \tensor{S}{^k_k} &\equiv& p_t^\text{eff} = \frac{1}{\eta} \Big\{ \left[ 4 \gamma - (d - 2) \right] p_t + \, 2 \gamma \left( \rho - p_r \right) - (d - 2) \tilde{\Lambda} \Big\}, \label{eq:Pteff}
\end{eqnarray}
where $k = 2, 3, \ldots, (d - 2)$. The radial and tangential pressures, $p_r$ and $p_t$, are equal due to the isotropic condition, hence $p_r = p_t = p$.

In this paper, we consider a spatially flat $d$-dimensional homogeneous and isotropic Friedmann-Lema\^{\i}tre-Robertson-Walker (FLRW) geometry, such that the interior region of the star is described by the metric
\begin{equation}
    ds^2 = -dt^2 + a^2(t) dr^2 + R^2(r, t) d\Omega^2_{d - 2}, \label{eq:FLRWmetric}
\end{equation}
where $a(t)$ is the scale factor, $R(r, t) = r a(t)$ is the physical radius of the collapsing body, and $d\Omega_{d - 2}^2$ is the line element on the unit $(d - 2)$-dimensional sphere. Using the metric above, Eq.~\eqref{eq:RastallIdea} takes the following form
\begin{eqnarray}
    && \left[ 1 - \frac{2 \gamma (d - 1)}{d - 2} \right] \dot{\rho} - \left[ \frac{2 \gamma (d - 1)}{d - 2} \right] \dot{p} + \, H (\rho + p) (d - 1) \left[ 1 - \frac{2 \gamma d}{d - 2} \right] = 0, \label{eq:ContinuityEquation}
\end{eqnarray}
and the $(00)$- and $(11)$-components of Eq.~\eqref{eq:RastallField} can be written respectively as
\begin{eqnarray}
    -2 \gamma (d - 1) \dot{H} - \frac{1}{2} (d - 1) \eta H^2 - \Lambda &=& \kappa \rho, \label{eq:FriedmannI} \\
    (\eta - 2 \gamma) \dot{H} + \frac{1}{2} (d - 1) \eta H^2 + \Lambda &=& \kappa p, \label{eq:FriedmannII}
\end{eqnarray}
where $H(t) = \dot{a}(t)/{a(t)}$ is the collapse rate and $\dot{x}$ denotes the derivative of a variable $x$ with respect to $t$. By adding the two equations above, we will get
\begin{equation}
    \dot{H} = -\frac{\kappa}{(d - 2)} (\rho + p). \label{eq:Hdot}
\end{equation}
Note that if we set $d = 4$ and $\Lambda = 0$, the field equations in Eqs.~\eqref{eq:FriedmannI} and \eqref{eq:FriedmannII} above will reduce to Eqs.~(8) and (9) in Ref.~\cite{Ziaie-et-al-gravitational-2019}, as expected.

In this model, we consider our matter as an isentropic perfect fluid, whose pressure is a linear function of the density. Hence, its equation of state will take the form $p = w \rho$, where $w$ is the barotropic index. By using Eqs.~\eqref{eq:FriedmannI} and \eqref{eq:FriedmannII}, the equation of state of the fluid then becomes
\begin{equation}
    \chi \dot{H} + \frac{1}{2} (1 + w) (d - 1) \eta H^2 + (1 + w) \Lambda = 0, \label{eq:DiffEqCollapseRate}
\end{equation}
with
\begin{equation}
    \chi \equiv 2 \gamma (1 + w) (d - 1) - (d - 2). \label{eq:chi}
\end{equation}

\section{The Case $\Lambda = 0$}

\label{sec-zero-Lambda}

Solving Eq.~\eqref{eq:DiffEqCollapseRate} above for the case of vanishing cosmological constant $(\Lambda = 0)$ and using the conditions $a(t_0) = a_0$ for a constant $a_0 > 0$ and $a(t_s) = 0$, where $[t_0, t_s]$ is the time interval of the collapse, will give us the scale factor,
\begin{equation}
    a(t) = a_0 \left( \frac{t_s - t}{t_s - t_0} \right)^\ell \label{eq:ScaleFactor}
\end{equation}
and the collapse rate,
\begin{equation}
    H(t) = -\frac{\ell}{t_s - t}, \label{eq:CollapseRate}
\end{equation}
where the constant $\ell$, which has an important role in the analysis later, is given by
\begin{equation}
    \ell \equiv \frac{2 \chi}{\eta (1 + w) (d - 1)}. \label{eq:ell}
\end{equation}
Note that the collapse rate $H(t)$ in Eq.~\eqref{eq:CollapseRate} is expected to be negative, hence $\ell > 0$.

We can differentiate Eq.~\eqref{eq:CollapseRate} with respect to $t$ and substitute it to Eq.~\eqref{eq:Hdot} to obtain the energy density,
\begin{equation}
    \rho(t) = \frac{(d - 2) \ell}{\kappa (1 + w)} \frac{1}{(t_s - t)^2}. \label{eq:EnergyDensity}
\end{equation}
Setting $t = t_0$, we find the expression for the collapse duration,
\begin{equation}
    t_s - t_0 = \sqrt{\frac{(d - 2) \ell}{\kappa (1 + w) \rho_0}}, \label{eq:collapse-duration-zero-Lambda}
\end{equation}
with $\rho_0 \equiv \rho(t_0)$. From Eq.~\eqref{eq:EnergyDensity} we can also see that the energy density blows up at $t = t_s$, which indicates the formation of spacetime singularity. This conclusion is supported by examining the Kretschmann scalar at $t = t_s$,
\begin{eqnarray}
    \mathcal{K} &=& \tensor{R}{_\mu_\nu_\alpha_\beta} \tensor{R}{^\mu^\nu^\alpha^\beta} \nonumber \\
    &=& 2 (d - 1) (2 \dot{H}^2 + d H^4 + 4 H^2 \dot{H}),
\end{eqnarray}
where $\tilde{R}$ and $\tilde{K}$ are the Ricci curvature and the Kretschmann scalar for $(d - 2)$-dimensional submanifold, respectively. Since $\mathcal{K} \sim (t - t_s)^{-4}$ diverges at $t = t_s$, it means that a spacetime singularity is formed at the end of the gravitational collapse.

Applying the equation of state $p = w \rho$ and the isotropic condition $p_r = p_t = p$, Eqs.~\eqref{eq:Rhoeff}-\eqref{eq:Pteff} become
\begin{eqnarray}
    \rho_\text{eff} &=& \frac{\chi}{\eta} \rho, \label{eq:rhoeff} \\
    p_{\text{eff}} &=& \left( -\frac{\chi}{\eta} + 1 + w \right) \rho, \label{eq:peff}
\end{eqnarray}
From these two expressions we note that the effective energy density and pressure satisfy
\begin{equation}
    \rho_\text{eff} + p_\text{eff} = (1 + w) \rho = \rho + p. \label{eq:RhoEffplusPEff}
\end{equation}
In order to ensure the physical reasonability condition for the collapsing scenario, we need to restrict the energy density to be nonnegative when measured by any nonlocal timelike observer. In other words, the energy-momentum tensor should satisfy the weak energy conditions (WEC) \cite{Joshi-Malafarina-recent-2011},
\begin{equation}
    \rho \geq 0, \qquad \rho + p \geq 0. \label{eq:WEC-zero}
\end{equation}
For our model, these conditions translate to
\begin{equation}
    \left. \begin{array}{cccc}
        & \rho_0 & \geq & 0 \\
        & w & \geq & -1 \\
        & \frac{\chi}{\eta} & \geq & 0
    \end{array} \right\}. \label{eq:WECresult}
\end{equation}
However, note that we always impose $w > -1$ to ensure that the collapse duration in Eq.~\eqref{eq:collapse-duration-zero-Lambda} remains finite.

\subsection{Singularity Formation}

\label{formation-singularity-zero-Lambda}

We will now discuss whether the outcome of the gravitational collapse in this model is a black hole, where the singularity is hidden behind the horizon, or a naked singularity, where the singularity can be seen by the external observers. The visibility of the singularity is determined by the structure of the trapped $(d - 2)$-submanifold during the collapse. This trapped submanifold is spacelike and topologically a sphere, such that both ingoing and outgoing null geodesics normal to it converge. (See Ref.~\cite{Senovilla-trapped-2011} for review.)

In order to describe these ingoing and outgoing null geodesics, let us introduce the null coordinates
\begin{eqnarray}
    d\xi^+ &=& \frac{1}{\sqrt{2}} \left[ dt - a(t) dr \right], \\
    d\xi^- &=& \frac{1}{\sqrt{2}} \left[ dt + a(t) dr \right],
\end{eqnarray}
so that our metric in Eq.~\eqref{eq:FLRWmetric} can be written in the double-null form as
\begin{equation}
    ds^2 = -2 d\xi^+ d\xi^- + R^2(r, t) d\Omega^2_{d - 2}, \label{eq:doublenullform}
\end{equation}
which then can be represented as
\begin{equation}
    \tilde{g}_{\mu \nu} = \begin{pmatrix} 0 & -1 & 0 \\ -1 & 0 & 0 \\ 0 & 0 & \sigma_{i j} \end{pmatrix}, \label{eq:newmetric}
\end{equation}
with $\sigma_{i j} = R^2(r, t) \hat{g}_{i j}$. In order to determine the trapped surfaces, we need to identify the expansion parameter of null geodesics, which is given by
\begin{equation}
    \Theta = \frac{1}{\sqrt{\sigma}} \frac{\partial}{\partial \lambda} \sqrt{\sigma},
\end{equation}
where $\sigma = \text{det}{(\sigma_{i j})}$. Therefore, we have
\begin{equation}
    \Theta = \frac{(d - 2)}{R(r, t)} \frac{\partial}{\partial \lambda} R(r, t),
\end{equation}
with two future oriented null geodesics, $\xi^{\pm} = \text{constant}$. Thus, by using $\lambda = \xi^\pm$, we will have
\begin{equation}
    \Theta_\pm = \frac{(d - 2)}{R(r, t)} \frac{\partial}{\partial \xi^\pm} R(r, t),
\end{equation}
with
\begin{eqnarray}
    \frac{\partial}{\partial \xi^+} &=& \frac{1}{\sqrt{2}} \left[ \frac{\partial}{\partial t} + \frac{1}{a(t)} \frac{\partial}{\partial r} \right], \nonumber \\
    \frac{\partial}{\partial \xi^-} & = & \frac{1}{\sqrt{2}} \left[ \frac{\partial}{\partial t} - \frac{1}{a(t)} \frac{\partial}{\partial r} \right]. \label{eq:partialxi}
\end{eqnarray}
Explicitly, we have
\begin{eqnarray}
    \Theta_\pm &=& \frac{d - 2}{\sqrt{2}} \left[ H(t) \pm \frac{1}{r a(t)} \right] \nonumber \\
    &=& \frac{d - 2}{\sqrt{2}} \left[ \frac{r \dot{a}(t) \pm 1}{r a(t)} \right]. \label{eq:Theta_pm}
\end{eqnarray}
If both $\Theta_+$ and $\Theta_-$ are positive or negative, then we will have a trapped spacetime. But if they have different signs, then we will have an untrapped spacetime. Therefore, we have the following conditions,
\begin{eqnarray}
    \Theta_+ \Theta_- &>& 0 \quad \text{trapped}, \label{eq:conditions1-1} \\
    \Theta_+ \Theta_- &<& 0 \quad \text{untrapped}, \label{eq:conditions1-2} \\
    \Theta_+ \Theta_- &=& 0 \quad \text{marginally trapped}. \label{eq:conditions1-3}
\end{eqnarray}
Since we are working with the gravitational collapse where $H(t) < 0$, then from Eq.~\eqref{eq:Theta_pm} we always have $\Theta_- < 0$, which means that the ingoing light rays are always going toward the center of the collapse. However, the sign of $\Theta_+$ depends on the value of $\ell$, which will be discussed later.

To determine the dynamics of the apparent horizon, let us consider the case where the $(d - 1)$-dimensional compact submanifold $\mathbb{S}^{d - 1}$ of radius $R(r, t)$ is spherically symmetric. The energy inside $\mathbb{S}^{d - 1}$, which is known as the Misner-Sharp energy \cite{Misner-Sharp-relativistic-1964,Bak-Rey-cosmic-2000}, has the form
\begin{equation}
    \mathcal{M} = \frac{(d - 1) (d - 2)}{16 \pi} v_{d - 1} R^{d - 3} \left[ 1 - h^{\mu \nu} (\partial_\mu R) (\partial_\nu R) \right], \label{eq:MisnerGeneral}
\end{equation}
where $v_{d - 1}$ is the volume of $\mathbb{S}^{d - 1}$ with unit radius. If $\mathbb{S}^{d - 1}$ is a $(d - 1)$-sphere, then $v_{d - 1}$ has the form
\begin{equation}
    v_{d - 1} = \frac{\pi^{(d - 1)/2}}{\Gamma{\left( \frac{d - 1}{2} + 1 \right)}},
\end{equation}
where $\Gamma(x)$ is the gamma function.

\begin{figure*}[t]
    \centering
    \includegraphics[width=\textwidth]{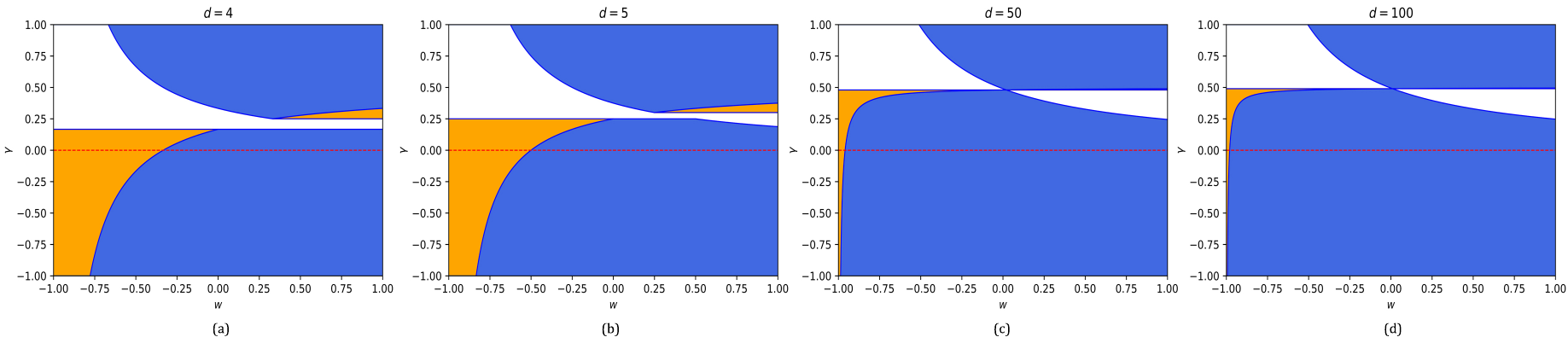}
    \caption{Possible final states of the gravitational collapse in Rastall gravity for various spacetime dimensions $d$ with and without cosmological constant. The general relativity limit, indicated by horizontal dashed line, is obtained by setting $\gamma = 0$. The blue (orange) region indicates the black hole (naked singularity) formation, while the white region is ruled out by the WEC. Note that for any finite dimension $d$ the naked singularity formation is possible, hence providing a counterexample to the CCC. However, in the limit of very high dimensions, the naked singularity regions in both the upper and lower branches are shrinking to zero, hence the CCC is fully restored in this limit. Note that this map also represents the case of nonvanishing cosmological constant, since the function $\ell$, which has a key role in determining the final outcome of the collapse, does not depend on the cosmological constant.}
    \label{fig-parameter-space}
\end{figure*}

Note that the term $\left[ 1 - h^{\mu \nu} (\partial_\mu R) (\partial_\nu R) \right]$ is a geometrical invariant \cite{Hayward-gravitational-1996}. Thus, we can calculate it in both coordinates $(t, r, \Omega_{d - 2})$ and $(\xi^+, \xi^-, \Omega_{d - 2})$,
\begin{eqnarray}
    \left[ h^{\mu \nu} (\partial_\mu R) (\partial_\nu R) \right] &=& 1 + \dot{R}^2 \\
    &=& \frac{2}{(d - 2)^2} R^2 \Theta_+ \Theta_-.
\end{eqnarray}
We then can write Eq.~\eqref{eq:MisnerGeneral} in both of these coordinates,
\begin{eqnarray}
    \mathcal{M} &=& \frac{(d - 1) (d - 2)}{16 \pi} v_{d - 1} R^{d - 3} \dot{R}^2, \label{eq:MisnerinCoordtr} \\
    &=& \frac{(d - 1) (d - 2)}{16 \pi} v_{d - 1} R^{d - 3} \left[ \frac{2}{(d - 2)^2} R^2 \Theta_+ \Theta_- - 1 \right]. \label{eq:MisnerinCoordxi}
\end{eqnarray}
The expansion parameter can be obtained by rearranging Eq.~\eqref{eq:MisnerinCoordxi} as
\begin{equation}
    \Theta_+ \Theta_- = \left[ \frac{16 \pi \mathcal{M}}{(d - 1) (d - 2) v_{d - 1} R^{d - 3}} - 1 \right] \frac{(d - 2)^2}{2 R^2}, \label{eq:ThetaMult}
\end{equation}
and the differentiation of Eq.~\eqref{eq:MisnerinCoordtr} with respect to $t$ and $r$ will give
\begin{eqnarray}
    \frac{\dot{\mathcal{M}}}{R^{d - 2} \dot{R}} &=& -\frac{(d - 1)}{8 \pi} v_{d - 1} \kappa p_\text{eff}, \\
    \frac{\mathcal{M}'}{R^{d - 2} R'} &=& \frac{(d - 1)}{8 \pi} v_{d - 1} \kappa \rho_\text{eff}, \label{eq:Misnerdiffr1}
\end{eqnarray}
where $x'$ denotes the derivative of a variable $x$ with respect to $r$. Another form of Misner mass can be obtained by integrating Eq.~\eqref{eq:Misnerdiffr1} with respect to $r$, such that we will have
\begin{equation}
    \mathcal{M} = \frac{v_{d - 1}}{8 \pi} \kappa \rho_\text{eff} R^{d - 1}. \label{eq:MisnerMassinrho}
\end{equation}
The expansion parameter can be recast in terms of the effective energy density by substituting Eq.~\eqref{eq:MisnerMassinrho} into \eqref{eq:ThetaMult}, such that we will obtain
\begin{equation}
    \Theta_+ \Theta_- = \left[\frac{2 \kappa \rho_\text{eff} R^2}{(d - 1) (d - 2)} - 1 \right] \frac{(d - 2)^2}{2 R^2}. \label{eq:ThetaMult2}
\end{equation}

Let us define a new function $\mathcal{A}$ as
\begin{equation}
    \mathcal{A} = \frac{2 \kappa \rho_\text{eff} R^2}{(d - 1) (d - 2)}. \label{eq:A0}
\end{equation}
Substituting Eqs.~\eqref{eq:ScaleFactor}, \eqref{eq:EnergyDensity}, and \eqref{eq:rhoeff} into Eq.~\eqref{eq:A0} will give us
\begin{equation}
    \mathcal{A}(t) = \mathcal{A}_0 \left(\frac{t_s - t}{t_s - t_0} \right)^{2 (\ell - 1)}, \label{eq:A-zeroLambda}
\end{equation}
with $\ell$ given in Eq.~\eqref{eq:ell}, and
\begin{equation}
    \mathcal{A}_0 = \frac{(1 + w) \ell}{d - 2} \kappa \rho_0 (r a_0)^2.
\end{equation}
Thus, the conditions \eqref{eq:conditions1-1}-\eqref{eq:conditions1-3} can be written as
\begin{eqnarray}
    \mathcal{A}(t) &>& 1 \quad \text{trapped}, \\
    \mathcal{A}(t) &<& 1 \quad \text{untrapped}, \\
    \mathcal{A}(t) &=& 1 \quad \text{marginally trapped}. \label{eq:conditions2}
\end{eqnarray}
Since the initial configuration of spacetime is not trapped, we have $\mathcal{A}_0 < 1$.

From Eq.~\eqref{eq:A-zeroLambda}, there are two cases for the outcomes of the gravitational collapse:

\textit{Case 1:} $\ell > 1$. The function $\mathcal{A}(t)$ in this case is monotonically decreasing to zero as the collapse occurs. Since $\mathcal{A}_0 < 1$, it means that $\mathcal{A}(t) < 1$ for all $t \in [t_0, t_s]$. Also, using Eqs.~\eqref{eq:Theta_pm} and \eqref{eq:ScaleFactor}, we can find that $\Theta_+ > 0$ for any $r < \infty$ as $t \to t_s$. Combined with $\Theta_- < 0$, it means that we will have an untrapped spacetime. In other words, there is no apparent horizon which can cover the singularity, since the outgoing radial null geodesics from the center of the collapse are untrapped. Therefore, it will form a naked singularity.

Furthermore, note that this naked singularity is timelike, ensuring that the outgoing radial light rays from the singularity can reach distant observers after the singularity is formed \cite{Goswami-et-al-timelike-2004}. To demonstrate this, let's define
\begin{eqnarray}
    R(t) = r a(t),
\end{eqnarray}
which then gives us
\begin{eqnarray}
    R' &=& \frac{dR}{dr} = a(t), \\
    \dot{R} &=& {dR}/{dt} = r \dot{a}(t).
\end{eqnarray}
Hence, we have ${R'}/{\dot{R}} = {dt}/{dr}$. At the singularity, we have $R(t) = 0$, and the metric is given by
\begin{eqnarray}
    ds^2 &=& -dt^2 + a^2(t) dr^2 \nonumber \\
    &=& \left[ -\left( \frac{R'}{\dot{R}} \right)^2 + a^2(t) \right] dr^2 \nonumber \\
    &=& \frac{a^2(t)}{r^2 \dot{a}^2(t)} \left[ \left( r \dot{a}(t) + 1 \right) \left( r \dot{a}(t) - 1 \right) \right] dr^2.
\end{eqnarray}
Since $\dot{a}(t) < 0$, we will find $\left( r \dot{a}(t) - 1 \right) < 0$. On the other hand, since $\left( r \dot{a}(t) + 1 \right)$ and $\Theta_+$ always have the same sign [see Eq.~\eqref{eq:Theta_pm}], we conclude that the sign of $ds^2$ is always opposite to the sign of $\Theta_+$. For the case of naked singularity, we have $\Theta_+ > 0$, such that $ds^2 < 0$, which means that the naked singularity is timelike.

\textit{Case 2:} $\ell < 1$. The function $\mathcal{A}(t)$ in this case is monotonically increasing as the collapse occurs, such that there exists a particular time $t = t_\text{AH}$ where $\mathcal{A}(t_\text{AH}) = 1$. Thus, an apparent horizon will form at $t = t_\text{AH}$ and the collapse continues until the singularity is formed at the end of the collapse. Again, from Eqs.~\eqref{eq:Theta_pm} and \eqref{eq:ScaleFactor}, we can find that $\Theta_+ < 0$ for any $r > 0$ as $t \to t_s$. Since $\Theta_- < 0$, it means that we will have a trapped spacetime. In other words, when the apparent horizon is formed, it will prevent all outgoing radial light rays to escape. Since $\Theta_+ < 0$, it implies that $ds^2 > 0$, which means that the singularity is spacelike. Therefore, the singularity will be covered and cannot be observed by external observers. In other words, it will form a black hole.

\begin{figure}[t]
    \centering
    \includegraphics[width=0.7\columnwidth]{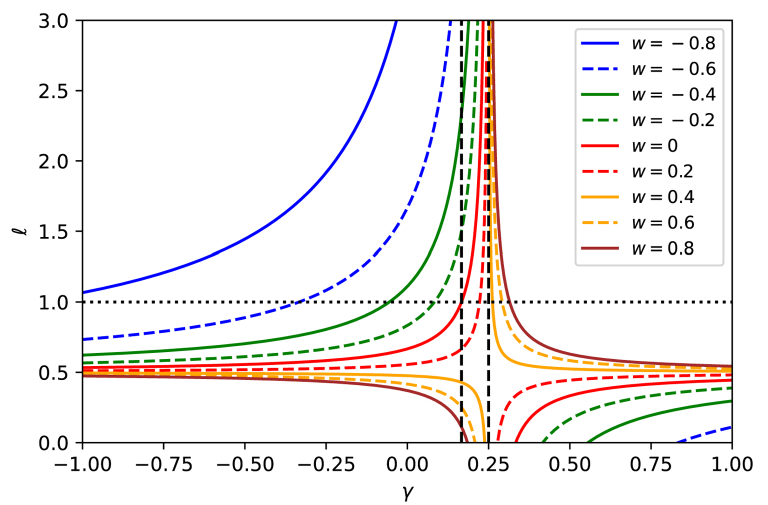}
    \caption{Plots of $\ell$ as a function of $\gamma$ for $d = 4$. If $\ell < 1$ ($\ell > 1$), the final state of the gravitational collapse is a black hole (naked singularity). Two dashed vertical lines represent the gap between the upper and lower branches in Fig.~\ref{fig-parameter-space}(a), which occurs in the interval $[\gamma_-, \gamma_+]$. Note that $\ell \to 2/d$ for $\gamma \to \pm \infty$ and for any $w \geq -1$. Therefore, naked singularity formation is possible only at values of $\gamma$ near the gap, while values of $\gamma$ sufficiently far from the gap always lead to black hole formation.}
    \label{fig-l-gamma}
\end{figure}

Following Ref.~\cite{Ziaie-et-al-gravitational-2019}, we map possible final states of the gravitational collapse in the case of $\Lambda = 0$ in the space of the Rastall parameter $\gamma$ and the barotropic index $w$ in Fig.~\ref{fig-parameter-space}(a-d). The blue region indicates the formation of black holes, the orange one indicates the formation of naked singularities, while the white one corresponds to the values of $(w, \gamma)$ which do not satisfy the WEC. The limit of Einstein gravity is obtained by setting $\gamma = 0$. Notice that there exists a gap in the interval $\gamma_- \leq \gamma \leq \gamma_+$ which separates the upper and lower branches, where
\begin{equation}
    \gamma_- = \frac{d - 3}{2 (d - 1)}, \qquad \gamma_+ = \frac{d - 2}{2 d}.
\end{equation}
The width of the gap, $\gamma_+ - \gamma_- = \frac{1}{d (d - 1)}$, approaches zero when $d \to \infty$, hence there is no `branch touching' for finite $d \geq 4$. Since $\ell \to 2/d$ for $\gamma \to \pm \infty$ and for any $w \geq -1$, which is always less than $1$ for $d \geq 4$, and since $\ell > 1$ always occurs near the gap (see Fig.~\ref{fig-l-gamma}), we conclude that naked singularity formation is possible only at the values of $\gamma$ near the gap and that values of $\gamma$ sufficiently far from the gap always lead to black hole formation.

Note that both naked singularity regions in the upper and lower branches of Fig.~\ref{fig-parameter-space} is shrinking when the spacetime dimension is increased. The naked singularity region in the lower branch will slowly turn to a black hole region, since the values of $\ell$ in this region, which are initially larger than unity, will become smaller than unity at higher dimensions. As for the naked singularity region in the upper branch, then note that its location extends from $w = 1/{(d - 1)}$ to higher values of $w$, and its top and bottom boundaries are given by
\begin{equation}
    \gamma_t = \frac{1}{2} - \frac{1}{(1 + w) (d - 1)}, \qquad \gamma_b = \frac{d - 2}{2 d},
\end{equation}
respectively. This region vanishes in the limit $d \to \infty$, since $\gamma_t, \gamma_b \to \frac{1}{2}$. Therefore, both naked singularity regions in the upper and lower branches are shrinking to zero when $d \to \infty$, hence we conclude that the CCC is fully restored in this limit, in agreement with the result in Ref.~\cite{Ghosh-Dadhich-naked-2001}.

Although we impose $w > -1$ for the rest of the paper, it is instructive to check the case where $w = -1$. Eq.~\eqref{eq:DiffEqCollapseRate} now becomes $-(d - 2) \dot{H} = 0$, which means that $a(t) = C_2 e^{C_1 t}$, where $C_1$ and $C_2$ are constants. Setting $a(t_0) = a_0$ and forcing the singularity to occur at $t = t_s$, namely $a(t_s) = 0$, we will have
\begin{equation}
    a(t) = a_0 e^{C_1 (t - t_0)}.
\end{equation}
The constant $C_1$ must be of negative value such that the singular collapse is possible, which occurs when $t_s - t_0 \to \infty$ such that the condition $a(t_s) = 0$ is satisfied. From this result we conclude that the collapse to singularity is possible for the case of $w = -1$, but the collapse has to occur in an infinite duration. This is in contrast with the case where $w > -1$, in which the collapse duration is finite (see Fig.~\ref{fig-DCollapse-d}). In order to fulfill the WEC, $\rho$ and $\rho_\text{eff}$ must be positive. Substituting $H = C_1$ and $\dot{H} = 0$ to Eq.~\eqref{eq:FriedmannI}, we will get
\begin{equation}
    \rho = -\frac{1}{2 \kappa} (d - 1) \eta C_1^2 - \frac{\Lambda}{\kappa}. \label{eq:rho_w=-1}
\end{equation}
The condition $\rho > 0$ then will give us a constraint to the value of $C_1$,
\begin{equation}
    C_1^2 < \frac{2}{d - 1} \frac{\Lambda}{\eta}.
\end{equation}
Since $\chi = -(d - 2)$ for the case of $w = -1$ [see Eq.~\eqref{eq:chi}], we can write Eq.~\eqref{eq:rhoeff} in this case as
\begin{equation}
    \rho_\text{eff} = -\frac{d - 2}{\eta} \rho. \label{eq:rhoeff_w=-1}
\end{equation}
The conditions $\rho > 0$ and $\rho_\text{eff} > 0$ then imply $\eta < 0$, which means that the collapse solution for the case $w = -1$ is possible only in the lower branch, in agreement with Fig.~\ref{fig-parameter-space}.

To check the endpoint of the collapse for the case $w = -1$, we need to calculate the function Eq.~\eqref{eq:A0}. From Eqs.~\eqref{eq:rho_w=-1} and \eqref{eq:rhoeff_w=-1} above, it is straightforward to prove that
\begin{equation}
    \mathcal{A}(t) = \mathcal{A}_0 e^{2 C_1 (t - t_0)}, \quad \mathcal{A}_0 = r^2 a_0^2 \left( C_1^2 + \frac{2}{d - 1} \frac{\Lambda}{\eta} \right).
\end{equation}
Since the initial configuration of spacetime is not trapped, we again should have $\mathcal{A}_0 < 1$. Therefore, since $C_1$ is of negative value, the function $\mathcal{A}(t)$ is monotonically decreasing when $t$ increases, hence the endpoint of the collapse in the case $w = -1$ will always be naked singularity. This is again in agreement with Fig.~\ref{fig-parameter-space}.

\subsection{Exterior Spacetime}

\label{exterior-spacetime-zero-Lambda}

Here we aim to find the spacetime metric for the exterior region outside the star and we require that both the interior and exterior spacetimes must be smooth at their mutual boundary $r = r_b$, the surface of the star. For the exterior region, we need to have a time-dependent generalization of the Schwarzschild metric which describes a black hole with a future event horizon. The most suitable choice is the Vaidya metric \cite{Vaidya-Shah-1957}, which can be generalized to higher dimensions as \cite{Ghosh-Dadhich-naked-2001}
\begin{equation}
    ds^2 = -f(r_v, V) dV^2 - 2 dr_v dV + r_v^2 d\Omega^2_{d - 2}, \label{eq:VaidyaMetric}
\end{equation}
where $f(r_v, V) = 1 - \frac{2 M(r_v, V)}{(d - 3) r_v^{d - 3}}$ is the exterior metric function, $M(r_v, V)$ is the mass enclosed by the $(d-2)$-sphere with the Vaidya radius $r_v$, and $V$ is the retarded null coordinate. Applying the Israel-Darmois junction conditions \cite{Israel-singular-1966}, we need to match the first and the second fundamental forms at the hypersurface $r = r_b$. The induced interior and exterior metrics in Eqs.~\eqref{eq:FLRWmetric} and \eqref{eq:VaidyaMetric}, which are also known as the first fundamental forms, can be written respectively as
\begin{eqnarray}
    ds_-^2 \big|_{r = r_b} &=& -dt^2 + r_b^2 a(t)^2 d\Omega^2_{d - 2}, \label{eq:interiormetric} \\
    ds_+^2 \big|_{r = r_b} &=& -\left[ f(r_v(t), V(t)) \dot{v}^2 + 2 \dot{r}_v \dot{V} \right] dt^2 + \, r_v^2(t) d\Omega^2_{d - 2}.
\end{eqnarray}
Matching both of these induced metrics will give us
\begin{equation}
    f(r_v(t), v(t)) \dot{V}^2 + 2 \dot{r}_v \dot{V} = 1, \qquad r_v = r_b a(t). \label{eq:FirstFormMatch}
\end{equation}

Note that the vector normal to the hypersurface of the interior metric is
\begin{equation}
    \mathbf{n}^i_- = \left[ 0, \frac{1}{a(t)}, 0, \ldots, 0 \right],
\end{equation}
while the nonzero components of the vector normal to the hypersurface of the exterior metric are
\begin{eqnarray}
    n^V_+ &=& -\frac{1}{\sqrt{f(r_v, V) + 2 \frac{dr_v}{dV}}}, \\
    n^{r_v}_+ &=& \frac{f(r_v, V) + \frac{dr_v}{dV}}{\sqrt{f(r_v, V) + 2 \frac{dr_v}{dV}}}.
\end{eqnarray}
The second fundamental form of the hypersurface, which is also known as the extrinsic curvature, can be calculated by taking the Lie derivative of the metric with respect to the normal vector $\bm{n}$,
\begin{eqnarray}
    K_{a b} &=& \frac{1}{2} \mathcal{L}_{\bm{n}} g_{a b} \nonumber \\
    &=& \frac{1}{2} \left[ g_{a b, c} n^c + g_{c b} n^c_{,a} + g_{a c} n^c_{,b} \right] .
\end{eqnarray}
The nonvanishing components of the extrinsic curvature of the hypersurface at $r = r_b$ for the interior and exterior regions are given by
\begin{eqnarray}
    \tensor{K}{^-_t_t} &=& 0, \\
    \tensor{K}{^-_{\theta_i}^{\theta_i}} &=& \frac{1}{r_b a(t)}, \\
    \tensor{K}{^+_t_t} &=& -\Bigg[ \dot{V}^2 \left( f f_{,r_v} \dot{V} + f_{,V} \dot{V} + 3 f_{,r_v} \dot{r}_v \right) + \, 2 \left( \dot{V} \ddot{r}_v - \dot{r}_v \ddot{V} \right) \Bigg] \nonumber \\
    && \times \, \frac{1}{2 \left( f \dot{V}^2 + 2 \dot{r}_v \dot{V} \right)^{\frac{3}{2}}}, \label{eq:Kttext} \\
    \tensor{K}{^+_{\theta_i}^{\theta_i}} &=& \frac{f \dot{V} + \dot{r}_v}{r_v \sqrt{f \dot{V}^2 + 2 \dot{r}_v \dot{V}}}.
\end{eqnarray}
Matching $\left[ \tensor{K}{^+_{\theta_i}^{\theta_i}} - \tensor{K}{^-_{\theta_i}^{\theta_i}} \right] \Big|_{r = r_b} = 0$ along with Eq.~\eqref{eq:FirstFormMatch} will give us
\begin{eqnarray}
    \dot{V} &=& \frac{1 + \sqrt{1 - f}}{f}, \label{eq:Vdot} \\
    \dot{r}_v &=& -\sqrt{1 - f}, \label{eq:Rdot}
\end{eqnarray}
where the minus sign in $\dot{r}_v$ indicates the collapse scenario. Matching $\left[ \tensor{K}{^+_t_t} - \tensor{K}{^-_t_t} \right] \Big|_{r = r_b} = 0$, along with Eqs.~\eqref{eq:Vdot} and \eqref{eq:Rdot} will give us
\begin{equation}
    f_{,V} = 0, \label{eq:fDiffwrtv}
\end{equation}
which shows that the exterior metric function $f(r_v, V)$, hence the mass function $M(r_v, V)$ also, must be independent of the retarded null coordinate $V$.

Next, we require the mass functions to be smooth at the boundary, namely $M(r_v) = \mathcal{M}(t, r_b)$. Hence, using Eq.~\eqref{eq:MisnerMassinrho} for $\mathcal{M}(t, r_b)$, we will have
\begin{eqnarray}
    \frac{2 M (r_v)}{(d - 3) r_v^{d - 3}} &=& \frac{v_{d - 1}}{8 \pi} \frac{(d - 1)}{(d - 3)} (1 + w) \ell \kappa \rho_0 (r_b a_0)^{2/\ell} r_v^{2 (\ell - 1)/\ell}.
\end{eqnarray}
Therefore, the exterior metric will take the form
\begin{equation}
    ds^2_+ = -\left( 1 - 2 M_0 r_v^{2 (\ell - 1)/\ell} \right) dV^2 - 2 dr_v dV + r_v^2 d\Omega^2_{d - 2},
\end{equation}
with $M_0$ is given by
\begin{equation}
    M_0 = \frac{v_{d - 1}}{16 \pi} \frac{(d - 1)}{(d - 3)} (1 + w) \ell \kappa \rho_0 (r_b a_0)^{2/\ell}.
\end{equation}

\section{The Case $\Lambda \neq 0$}

\label{sec-nonzero-Lambda}

If the cosmological constant is nonzero, $\Lambda \neq 0$, solving Eq.~\eqref{eq:DiffEqCollapseRate} will give us two types of general solutions: trigonometric and exponential forms. The former (latter) can be obtained by setting $\Lambda$ and $\eta$ to have the same sign (different signs). Note that the sign of $\eta$ is positive (negative) for the upper (lower) branch in the parameter space in Fig.~\ref{fig-parameter-space}, so these two branches can be identified using the sign of $\eta$. Furthermore, since $\ell \sim \chi/\eta$ is always positive (see below), it means that $\chi$ and $\eta$ always have the same sign in both solutions.

Using the boundary conditions $a(t_0) = a_0$ and $a(t_s) = 0$, we can find the scale factors as
\begin{eqnarray}
    a_\text{trig}(t) &=& a_0 \left[ \frac{\sin{[K_\text{trig} (t_{s, \text{trig}} - t)]}}{\sin{[K_\text{trig} (t_{s, \text{trig}} - t_0)]}} \right]^\ell, \label{eq:ScaleFactorCosmoTrig} \\
    a_\text{exp}(t) &=& a_0 e^{-K_\text{exp} \ell \left( t - t_0 \right)} \left[ \frac{1 - e^{-2 K_\text{exp} \left( t_{s, \text{exp}} - t \right)}}{1 - e^{-2 K_\text{exp} \left( t_{s, \text{exp}} - t_0 \right)}} \right]^\ell, \label{eq:ScaleFactorCosmoExp}
\end{eqnarray}
and the collapse rates as
\begin{eqnarray}
    H_\text{trig}(t) &=& -\frac{K_\text{trig} \ell}{\tan{[K_\text{trig} (t_{s, \text{trig}} - t)]}}, \label{eq:CollapseRateCosmoTrig} \\
    H_\text{exp}(t) &=& -K_\text{exp} \ell \left[ \frac{1 + e^{-2 K_\text{exp} (t_{s, \text{exp}} - t)}}{1 - e^{-2 K_\text{exp} (t_{s, \text{exp}} - t)}} \right], \label{eq:CollapseRateCosmoExp}
\end{eqnarray}
where the constants $K_\text{trig}$ and $K_\text{exp}$ are given by
\begin{eqnarray}
    K_\text{trig} &\equiv& \sqrt{\frac{\left( 1 + w \right)}{\ell} \frac{\Lambda}{\chi}}, \label{eq:Ktrig} \\
    K_\text{exp} &\equiv& \sqrt{-\frac{\left( 1 + w \right)}{\ell} \frac{\Lambda}{\chi}}. \label{eq:Kexp}
\end{eqnarray}
Note that the subscripts `trig' and `exp' indicate the trigonometric and exponential solutions, respectively.

From Eqs.~\eqref{eq:Hdot}, \eqref{eq:CollapseRateCosmoTrig}, and \eqref{eq:CollapseRateCosmoExp}, we can get the expressions for the energy density,
\begin{eqnarray}
    \rho_\text{trig}(t) &=& \rho_{0, \text{trig}} \left[ \frac{\sin{\left[ K_\text{trig} \left( t_{s, \text{trig}} - t_0 \right) \right]}}{\sin{\left[ K_\text{trig} \left( t_{s, \text{trig}} - t \right) \right]}} \right]^2, \label{eq:EnergyDensityCosmoTrig} \\
    \rho_\text{exp}(t) &=& \rho_{0, \text{exp}} e^{-2 K_\text{exp} \left( t_0 - t \right)} \left[ \frac{1 - e^{-2 K_\text{exp} \left( t_{s, \text{exp}} - t_0 \right)}}{1 - e^{-2 K_\text{exp} \left( t_{s, \text{exp}} - t \right)}} \right]^2, \label{eq:EnergyDensityCosmoExp}
\end{eqnarray}
where
\begin{eqnarray}
    \rho_{0, \text{trig}} &=& \frac{\Lambda}{\chi} \frac{\left( d - 2 \right)}{\kappa} \frac{1}{\sin^2{\left[ K_\text{trig} \left( t_{s, \text{trig}} - t_0 \right) \right]}}, \label{eq:InitialEnergyDensityCosmoTrig} \\
    \rho_{0, \text{exp}} &=& 4 \left| \frac{\Lambda}{\chi} \right| \frac{\left( d - 2 \right)}{\kappa} \frac{e^{-2 K_\text{exp} \left( t_{s, \text{exp}} - t_0 \right)}}{\left[ 1 - e^{-2 K_\text{exp} \left( t_{s, \text{exp}} - t_0 \right)} \right]^2}. \label{eq:InitialEnergyDensityCosmoExp}
\end{eqnarray}

Given the initial values for $\rho_{0, \text{trig}}$ and $\rho_{0, \text{exp}}$, we can invert the last two equations to obtain the collapse durations for both solutions,
\begin{eqnarray}
    t_{s, \text{trig}} - t_0 &=& \frac{1}{K_\text{trig}} \arctan{\left[ \frac{1}{\zeta_{\text{trig}}} \right]}, \label{eq:DurationCollapseTrig}\\
    t_{s, \text{exp}} - t_0 &=& \frac{1}{2 K_\text{exp}} \ln{\left( \frac{\zeta_{\text{exp}} + 1}{\zeta_{\text{exp}} - 1} \right)} \label{eq:DurationCollapseExp},
\end{eqnarray}
where
\begin{eqnarray}
    \zeta_{\text{trig}} &\equiv& \sqrt{\frac{\kappa \rho_{0, \text{trig}}}{\left( d - 2 \right)} \frac{\chi}{\Lambda} - 1}, \label{eq:zeta_trig} \\
    \zeta_{\text{exp}} &\equiv& \sqrt{-\frac{\kappa \rho_{0, \text{exp}}}{\left( d - 2 \right)} \frac{\chi}{\Lambda} + 1}. \label{eq:zeta_exp}
\end{eqnarray}
Note that from Eqs.~\eqref{eq:EnergyDensityCosmoTrig} and \eqref{eq:EnergyDensityCosmoExp}, we can see that the energy densities $\rho_\text{trig}(t)$ and $\rho_\text{exp}(t)$ blow up at $t = t_s$. Again, together with the divergence of the Kretschmann scalars at $t = t_s$, namely $\mathcal{K} \sim \sin^{-4}{[K (t_s - t)]}$ for the trigonometric solution and $\mathcal{K} \sim \left[ 1 - e^{-2 K (t_s - t)} \right]^{-4}$ for the exponential solution, we conclude that the spacetime singularity is formed at $t = t_s$.

\begin{figure}[t]
    \centering
    \includegraphics[width=0.7\columnwidth]{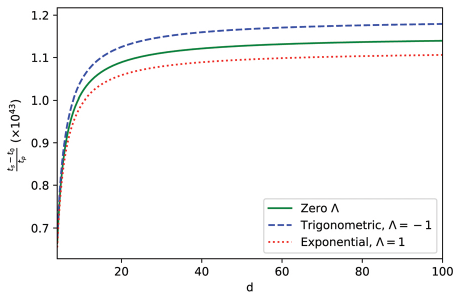}
    \caption{Plots of the collapse duration $t_s - t_0$ as a function of $d$ for $\gamma =  0$, $w = 0$ and $\Lambda = \pm 1$ in the exponential (trigonometric) case. Here we have rescaled the vertical axis using the Planck time $t_P$. Note that the trigonometric (exponential) solution always yields larger (smaller) collapse duration than the case of zero cosmological constant. This is because the effective energy density for the trigonometric (exponential) case is smaller (larger) than the one for the zero cosmological constant case.}
    \label{fig-DCollapse-d}
\end{figure}

From Eqs.~\eqref{eq:Rhoeff}-\eqref{eq:Pteff}, we can obtain the expressions for the effective energy density and pressure as
\begin{eqnarray}
    \rho_\text{eff} &=& \frac{\chi}{\eta} \rho - (d - 2) \frac{\tilde{\Lambda}}{\eta}, \label{eq:rhoeffCosmo} \\
    p_\text{eff} &=& \frac{\eta (1 + w) - \chi}{\eta} \rho + (d - 2) \frac{\tilde{\Lambda}}{\eta}. \label{eq:peffCosmo}
\end{eqnarray}
Note that in this case the effective energy density and pressure still satisfy
\begin{eqnarray}
    \rho_\text{eff} + p_\text{eff} = (1 + w) \rho = \rho + p. \label{eq:RhoEffplusPEffCosmo}
\end{eqnarray}
Moreover, the WEC must again be satisfied in order to ensure that the collapsing scenario is physically reasonable,
\begin{equation}
    \rho \geq 0, \qquad \rho + p \geq 0. \label{eq:WEC-nonzero}
\end{equation}
For the trigonometric case, these conditions then translate to
\begin{equation}
    \left. \begin{array}{cccc}
        & \frac{\Lambda}{\chi} & \geq & 0 \\
        & w & \geq & -1 \\
        & \frac{\Lambda}{\eta} & \geq & 0
    \end{array} \right\}, \label{eq:WECTrig}
\end{equation}
while for the exponential case they become
\begin{equation}
    \left. \begin{array}{cccc}
        & - \frac{\Lambda}{\chi } & \geq & 0 \\
        & w & \geq & -1 \\
        & - \frac{\Lambda}{\eta} & \geq & 0
    \end{array} \right\}. \label{eq:WECExp}
\end{equation}
Similar with the case $\Lambda = 0$, we always impose $w > -1$ to ensure that $K_\text{trig}$ and $K_\text{exp}$ in Eqs.~\eqref{eq:Ktrig} and \eqref{eq:Kexp} are strictly positive, such that the collapse durations in Eqs.~\eqref{eq:DurationCollapseTrig} and \eqref{eq:DurationCollapseExp} are finite. Therefore, if we restrict $t_{s, \text{trig}} - t$ in the trigonometric case to satisfy
\begin{equation}
    t_{s, \text{trig}} - t_0 \leq \frac{\pi}{2 K_\text{trig}} \label{eq:ts_trig_limitation}
\end{equation}
in order to avoid singularity formation before $t = t_{s, \text{trig}}$, the value of $\ell$ must be strictly positive to ensure that the collapse rates in Eqs.~\eqref{eq:CollapseRateCosmoTrig} and \eqref{eq:CollapseRateCosmoExp} are negative.

Writing Eq.~\eqref{eq:rhoeffCosmo} for both trigonometric and exponential solutions,
\begin{eqnarray}
    \rho_{\text{eff}_\text{trig}} &=& \frac{\chi}{\eta} \rho - (d - 2) \left| \frac{\tilde{\Lambda}}{\eta} \right|, \label{eq:rhoeffCosmoTrig} \\
    \rho_{\text{eff}_\text{exp}} &=& \frac{\chi}{\eta} \rho + (d - 2) \left| \frac{\tilde{\Lambda}}{\eta} \right|, \label{eq:rhoeffCosmoExp}
\end{eqnarray}
and also from Eq.~\eqref{eq:rhoeff} for the case of zero cosmological constant, we find that
\begin{equation}
    \rho_{\text{eff}_\text{trig}} < \rho_{\text{eff}_{\Lambda = 0}} < \rho_{\text{eff}_\text{exp}}.
\end{equation}
Since higher $\rho_\text{eff}$ will give us a stronger gravitational attraction in the collapse process, it means that the collapse duration is smaller when $\rho_\text{eff}$ is higher, which then explains intuitively why the trigonometric (exponential) case has a slower (faster) collapse than the zero cosmological constant case for any spacetime dimension (see Fig.~\ref{fig-DCollapse-d}).

\begin{figure}[t]
    \centering
    \includegraphics[width=0.7\columnwidth]{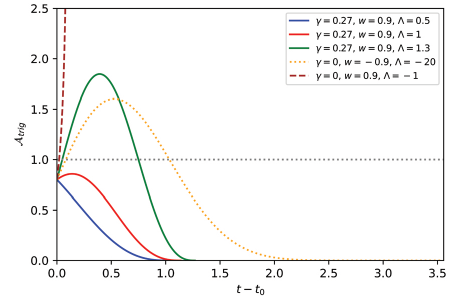}
    \caption{Plots of $\mathcal{A}_\text{trig}$ as a function of $t - t_0$ at $d = 4$. We set $\mathcal{A}_{0, \text{trig}} = 0.8$, so that the spacetime is initially untrapped. For $\gamma = 0$, $w = 0.9$ (dashed red line), which corresponds to a point in the blue area in Fig.~\ref{fig-parameter-space}, the value of $\mathcal{A}_\text{trig}$ becomes greater than unity as the time elapses, hence leading to the formation of a black hole. For $\gamma = 0.27$, $w = 0.9$ (blue, red, and green solid lines) and $\gamma = 0$, $w = -0.9$ (dotted orange line), which correspond to points in the naked singularity regions in the upper and lower branches of Fig.~\ref{fig-parameter-space}, respectively, the values of $\mathcal{A}_\text{trig}$ approach zero as the time elapses, indicating that the final states will be naked singularities. However, for the naked singularity region in the upper (lower) branch, there exists a critical value $\Lambda_\text{crit}$ such that if $\Lambda > \Lambda_\text{crit} > 0$ ($\Lambda < \Lambda_\text{crit} < 0$) and if $\Lambda$ does not go beyond a cutoff value $\Lambda_\text{cut}$ (see Fig.~\ref{fig-DCollapse-Lambda}), there will be a time period where $\mathcal{A}_\text{trig}$ is larger than unity before going back to less than unity afterwards (see the solid green and the dotted orange lines in the figure above), indicating the emergence of a temporary trapped surface. When the trapped surface is formed, the collapse may be seen to form a black hole by distant observers with observational timescale much shorter than the collapse duration. However, this trapped surface is formed only temporarily such that at later times the singularity will eventually be seen as a naked singularity by distant observers with observational timescale comparable to the collapse duration.}
    \label{fig-Atrig}
\end{figure}

\begin{figure*}[t]
    \centering
    \includegraphics[width=\textwidth]{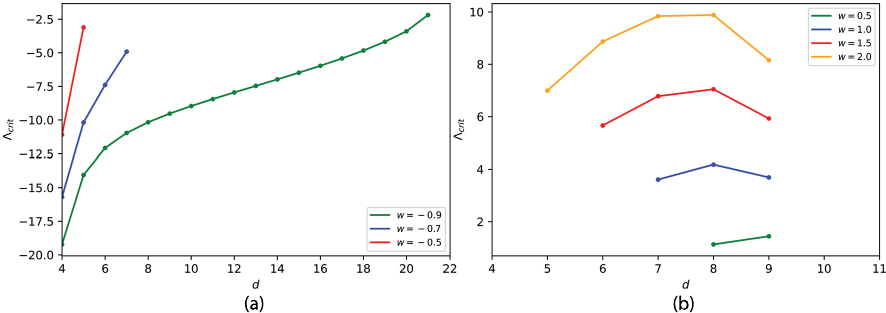}
    \caption{Plots of the critical values $\Lambda_\text{crit}$ that could lead to the emergence of temporary trapped surfaces for several values of $w$. Here we assume $\mathcal{A}_{0_{trig}} = 0.8$. For $\gamma = 0$ ($\gamma = 0.4$) in Subfigure a (b), which corresponds to the naked singularity region in the lower (upper) branch of Fig.~\ref{fig-parameter-space}, $\Lambda_\text{crit}$ is negative (positive), which implies that any $\Lambda < \Lambda_\text{crit} < 0$ ($\Lambda > \Lambda_\text{crit} > 0$) could lead to the emergence of temporary trapped surfaces, as long as $\Lambda$ does not go beyond the cutoff value $\Lambda_\text{cut}$ (see Fig.~\ref{fig-DCollapse-Lambda}).}
    \label{fig-LambdaTemp-d}
\end{figure*}

\subsection{Singularity Formation}

\label{singularity-formation-nonzero-Lambda}

In this Subsection, we use the same procedure as in Subsection \ref{formation-singularity-zero-Lambda}. Substituting Eqs.~\eqref{eq:ScaleFactorCosmoTrig}, \eqref{eq:ScaleFactorCosmoExp}, \eqref{eq:EnergyDensityCosmoTrig}, \eqref{eq:EnergyDensityCosmoExp} and \eqref{eq:rhoeffCosmo} to Eq.~\eqref{eq:A0}, we will obtain
\begin{eqnarray}
    \mathcal{A}_{\text{trig}}(t) &=& \mathcal{A}_{0, \text{trig}} \left[ \frac{\cos{[K_{\text{trig}} (t_{s, \text{trig}} - t)]}}{\cos{[K_{\text{trig}} (t_{s, \text{trig}} - t_0)]}} \right]^2 \left[ \frac{\sin{[K_\text{trig} (t_{s, \text{trig}} - t)]}}{\sin{[K_\text{trig} (t_{s, \text{trig}} - t_0)]}} \right]^{2 (\ell - 1)}, \\
    \mathcal{A}_{\text{exp}}(t) &=& \mathcal{A}_{0, \text{exp}} e^{-2 K_\text{exp} \ell \left(t - t_0\right)} \left[ \frac{1 + e^{-2 K_\text{exp} \left(t_{s, \text{exp}} - t\right)}}{1 + e^{-2 K_\text{exp} \left(t_{s, \text{exp}} - t_0\right)}} \right]^2 \nonumber \\
    && \times \left[ \frac{1 - e^{-2 K_\text{exp} \left(t_{s, \text{exp}} - t\right)}}{1 - e^{-2 K_\text{exp} \left(t_{s, \text{exp}} - t_0\right)}} \right]^{2 (\ell - 1)},
\end{eqnarray}
with
\begin{eqnarray}
    \mathcal{A}_{0, \text{trig}} &=& \frac{2 (r a_0)^2}{(d - 1)} \frac{\Lambda}{\eta} \zeta_{\text{trig}}^2, \\
    \mathcal{A}_{0, \text{exp}} &=& \frac{2 (r a_0)^2 }{(d - 1)} \left| \frac{\Lambda}{\eta} \right| \zeta_{\text{exp}}^2.
\end{eqnarray}
Since there is no trapped surface in the initial configuration before the collapse, we set $\mathcal{A}_{0, \text{trig}} < 1$ and $\mathcal{A}_{0, \text{exp}} < 1$.

There are two cases here for the outcomes of the gravitational collapse:

\textit{Case 1:} $\ell < 1$. There exist particular times $t = t_{\text{AH}_\text{trig}}$ and $t = t_{\text{AH}_\text{exp}}$ such that $\mathcal{A}_{\text{trig}}(t) \geq 1$ and $\mathcal{A}_{\text{exp}}(t) \geq 1$ for $t \geq t_{\text{AH}_\text{trig}}$ and $t \geq t_{\text{AH}_\text{exp}}$, respectively. Hence, apparent horizon will form at $t = t_{\text{AH}_\text{trig}}$ and $t = t_{\text{AH}_\text{exp}}$ before the singularity formation at the end of the collapse. Therefore, the final state of the collapse will be a black hole.

\textit{Case 2:} $\ell > 1$. Here we have
\begin{eqnarray}
    \mathcal{A}_{\text{trig}}(t \to t_{s, \text{trig}}) &\to& 0, \\
    \mathcal{A}_{\text{exp}}(t \to t_{s, \text{exp}}) &\to& 0,
\end{eqnarray}
at the end of the collapse, so that the singularity will be visible to outside observers, leading to the naked singularity formation.

The map for possible final states of the gravitational collapse in the case of $\Lambda \neq 0$ is identical with Fig.~\ref{fig-parameter-space}. In other words, the final outcome of the collapse is determined only by the value of $\ell$, which does not depend on $\Lambda$. However, the relative signs between $\Lambda$ and $\eta$ will determine the form of the solution for the collapse evolution. Since the upper branch in Fig.~\ref{fig-parameter-space} has positive $\eta$, then the black hole and naked singularity regions in the upper branch will have trigonometric (exponential) solutions if $\Lambda$ is positive (negative). On the other hand, since the lower branch has negative $\eta$, then the black hole and naked singularity regions in the lower branch will have trigonometric (exponential) solutions if $\Lambda$ is negative (positive).

\begin{figure*}[t]
    \centering
    \includegraphics[width=\textwidth]{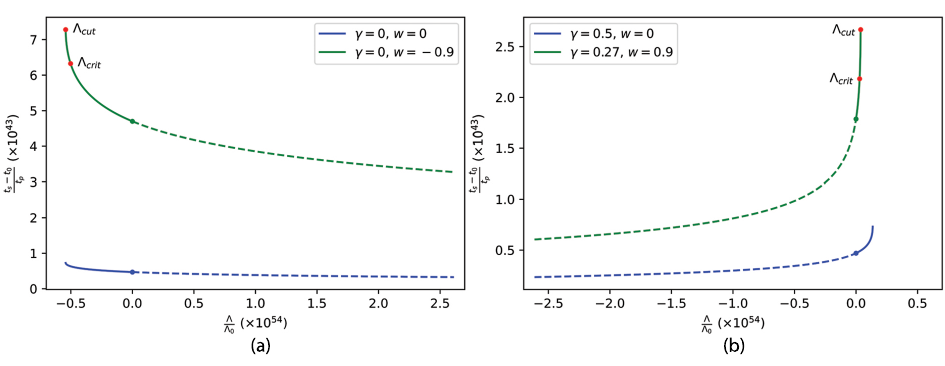}
    \caption{Plots of the collapse duration $t_s - t_0$ as a function of $\frac{\Lambda}{\Lambda_0}$, with $\Lambda_0 = 10^{-52}$ $\text{m}^{-2}$, at $d = 4$ for exponential (dashed) and trigonometric (solid) solutions for values of $(w, \gamma)$ corresponding to points in the lower (Subfigure a) and upper (Subfigure b) branches of Fig.~\ref{fig-parameter-space}. The blue (green) lines indicate that the endpoint of the collapse is a black hole (naked singularity). Note that there exists a cutoff value $\Lambda_\text{cut} < 0$ ($\Lambda_\text{cut} > 0$) of the cosmological constant for the trigonometric solution in the lower (upper) branch such that there is no possible collapse if $\Lambda < \Lambda_\text{cut}$ ($\Lambda > \Lambda_\text{cut}$). However, for the trigonometric solution in the naked singularity region only, there also exists a critical value $\Lambda_\text{crit}$ of the cosmological constant at the interval $[\Lambda_\text{cut}, 0]$ ($[0, \Lambda_\text{cut}]$) in the naked singularity region in the lower (upper) branch such that temporary trapped surfaces emerge when $\Lambda_\text{cut} < \Lambda < \Lambda_\text{crit} < 0$ ($0 < \Lambda_\text{crit} < \Lambda < \Lambda_\text{cut}$).}
    \label{fig-DCollapse-Lambda}
\end{figure*}

Although the possible final states of the collapse in both cases of zero and nonzero cosmological constants are described in the same Fig.~\ref{fig-parameter-space}, there is a striking difference between these two cases especially when we focus on the trigonometric solutions at the naked singularity regions in both upper and lower branches. For the case of $\Lambda = 0$, the value of $\mathcal{A}(t)$ is always monotonically decreasing to zero [see Eq.~\eqref{eq:A-zeroLambda}]. However, for $\Lambda \neq 0$, especially in the case of trigonometric solution, the value of $\mathcal{A}_{\text{trig}}(t)$ can be nonmonotonic. Therefore, there exists a critical value $\Lambda_\text{crit}$, which takes a negative (positive) value for the lower (upper) naked singularity region, such that if $\Lambda < \Lambda_\text{crit} < 0$ ($\Lambda > \Lambda_\text{crit} > 0$) then there will be a time period $T$ in which $\mathcal{A}_{\text{trig}}(t)$ may become larger than unity before going back to less than unity afterwards (see Fig.~\ref{fig-Atrig}). This indicates a possible emergence of temporary trapped surface, where the collapsing object will not be visible to distant observers in that period $T$. The critical values $\Lambda_\text{crit}$ for the naked singularity regions in both the upper and lower branches for various values of parameters $(w, \gamma)$ are plotted in Fig.~\ref{fig-LambdaTemp-d}.

Note that the term ``temporary trapped surface'' that we use here has a different context with the one discussed in the literature, for example in Ref.~\cite{Bambi-et-al-black-2014,Bambi-et-al-terminating-2014,Mersini-Houghton-backreaction-2014}. In these works, the collapse process may not form a singularity but instead only creates a temporary trapped surface. In other words, the singularity is replaced by a bounce, such that the scale factor $a(t)$ only has a nonzero minimum but never goes to zero. However, in our work, we always assume that $a(t_s) = 0$, hence the collapse discussed in this paper always forms a singularity as the endpoint. Therefore, when the temporary trapped surface is formed, the collapse may be seen to form a black hole by distant observers with observational timescale much shorter than the collapse duration $t_s - t_0$. However, this trapped surface is formed only temporarily such that at later times the singularity will eventually be seen as a naked singularity by distant observers which have an observational timescale comparable to the collapse duration. Hence, according to some distant observers with observational timescale much shorter than the collapse duration, the naked singularity regions in Fig.~\ref{fig-parameter-space} will look possibly narrower than what are depicted, since some parts of these regions may be observed as black holes.

Note that the collapse duration depends strongly on the magnitude of the cosmological constant (see Fig.~\ref{fig-DCollapse-Lambda}). Notice the striking difference in the behavior of the trigonometric and exponential solutions as $\Lambda$ varies: the collapse duration for the exponential solution decreases and asymptotically approaches zero as $| \Lambda | \to \infty$, but it increases in the trigonometric case as $| \Lambda |$ increases, until it reaches a cutoff value $| \Lambda_\text{cut} |$. Note that $\Lambda_\text{cut}$ is negative (positive) for the lower (upper) branch. The behavior of the trigonometric and exponential solutions in Fig.~\ref{fig-DCollapse-Lambda} is in agreement with Fig.~\ref{fig-DCollapse-d}, where the collapse duration for the trigonometric (exponential) solution is larger (smaller) than the collapse duration for the solution with zero cosmological constant. Also, from Eq.~\eqref{eq:rhoeffCosmoTrig} above, increasing $| \tilde{\Lambda} |$ in the trigonometric solution will decrease the effective energy density of the star $\rho_{\text{eff}_\text{trig}}$, hence the collapse duration will increase as $| \tilde{\Lambda} |$ increases. However, the collapse will not be possible anymore when the value of $| \tilde{\Lambda} |$ reaches the cutoff value $| \tilde{\Lambda}_\text{cut} |$ in which $\rho_{\text{eff}_\text{trig}}$ vanishes. Thus, for the trigonometric solution, the minimum value for the energy density that leads to a singular collapse scenario is
\begin{equation}
    \rho_{\text{trig}_\text{min}} = (d - 2) \left| \frac{\tilde{\Lambda}_\text{cut}}{\chi} \right|.
\end{equation}
As for the exponential solution, we see from Eq.~\eqref{eq:rhoeffCosmoExp} that there is no restriction for the value of $| \tilde{\Lambda} |$, hence there is no cutoff value $| \tilde{\Lambda}_\text{cut} |$ in this case. Furthermore, since increasing $| \tilde{\Lambda} |$ will increase $\rho_{\text{eff}_\text{exp}}$, the collapse duration will decrease as $| \tilde{\Lambda} |$ increases. Note that the emergence of temporary trapped surface, which can occur only in the case of trigonometric solution in the naked singularity region, together with the fact that the trigonometric and exponential solutions give us different collapse durations as in Fig.~\ref{fig-DCollapse-d}, indicates that these two solutions are physically distinct.

\subsection{Exterior Spacetime}

\label{exterior-spacetime-nonzero-Lambda}

\begin{figure*}[t]
    \centering
    \includegraphics[width=\textwidth]{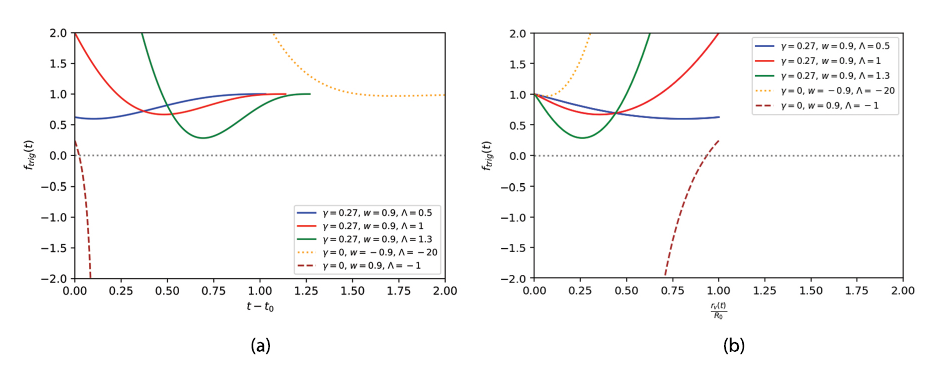}
    \caption{Plots of $f_\text{trig}$ at $d = 4$: (a) as a function of $t - t_0$ and (b) as a function of $r_v$ through the parameter $t$. Here we also set $\mathcal{A}_{0, \text{trig}} = 0.8$. Note that $R_0 = r_b a_0$ is the initial radius of the star. When the final state of the gravitational collapse is a black hole (dashed red line), the horizon in the external spacetime is formed. However, when the final state of the collapse is a naked singularity (all solid lines and dotted orange line), no horizon forms in the exterior spacetime. Therefore, when the temporary trapping occurs during the gravitational collapse to naked singularity (which occurs in the case of the solid green and the dotted orange lines, see Fig.~\ref{fig-Atrig}), the inability of light rays to reach distant observers is not due to the presence of an event horizon in the exterior spacetime but rather results from the complex collapse dynamics occurring within the interior spacetime.}
    \label{fig-ftrig}
\end{figure*}

To describe the exterior spacetime, we will use the higher-dimensional Vaidya-(anti-)de Sitter metric,
\begin{equation}
    ds^2 = -f(r_v, V) dV^2 - 2 dr_v dV + r_v^2 d\Omega^2_{d - 2}, \label{eq:VaidyaMetric1}
\end{equation}
where $f(r_v, V)$ is the exterior metric function, which contains the cosmological constant term,
\begin{equation}
    f(r_v, V) = 1 - 2 \frac{M(r_v, V)}{(d - 3) r_v^{(d - 3)}} + \frac{2 \Lambda}{\eta (d - 1)} r_v^2, \label{eq:VaidyaMetricNonzeroLambda}
\end{equation}
$M(r_v, V)$ is the mass enclosed by the $(d - 2)$-sphere with the Vaidya radius $r_v$, and $V$ is the retarded null coordinate. Note that if we set $d = 4$ in the general relativity limit $\gamma = 0$ such that $\eta = -2$, this metric will be identical with the one in Ref.~\cite{Wagh-Maharaj-naked-1999}.

Matching the first and the second fundamental forms will give us the same result as in the case of zero cosmological constant, since the derivation to prove this result in Eqs.~\eqref{eq:interiormetric}-\eqref{eq:fDiffwrtv} is general, without assuming any specific form for the function $f(r_v, V)$ such as Eq.~\eqref{eq:VaidyaMetric}, hence it can be applied for the case of $\Lambda \neq 0$ as well. Therefore, since $f_{,V} = 0$, it again implies that the exterior metric function $f(r_v, V)$ and also $M(r_v, V)$ in Eq.~\eqref{eq:VaidyaMetricNonzeroLambda} must be independent of the retarded null coordinate $V$.

Matching the mass functions $M(r_v) = \mathcal{M}(t, r_b)$ for both trigonometric and exponential solutions, and using Eq.~\eqref{eq:MisnerMassinrho} for $\mathcal{M}(t, r_b)$, we will have
\begin{eqnarray}
    \frac{2 M_\text{trig(exp)}(r_v)}{(d - 3) r_v^{(d - 3)}} &=& -\frac{v_{(d - 1)}}{8 \pi} \frac{(d - 1) (d - 2)}{(d - 3)} \frac{\Lambda}{\chi} (1 + w) \ell r_v^2 \nonumber \\
    && + \, \frac{v_{(d - 1)}}{8 \pi} \frac{(d - 1)}{(d - 3)} (1 + w) \ell \kappa \rho_{0, \text{trig(exp)}} (r_b a_0)^{2/\ell} r_v^{{2 (\ell - 1)}/\ell}.
\end{eqnarray}
Therefore, the exterior metrics for both trigonometric and exponential solutions are given as
\begin{eqnarray}
    && ds^2_{+, \text{trig(exp)}} = -f_\text{trig(exp)}(r_v) dV^2 - 2 dr_v dV + r_v^2 d\Omega^2_{d - 2},\label{eq:VaidyaMetric2}
\end{eqnarray}
where
\begin{eqnarray}
    f_\text{trig(exp)}(r_v) &=& 1 - 2 M_{0, \text{trig(exp)}} r_v^{2 (\ell - 1)/\ell} \nonumber \\
    && + \, \Bigg( 1 + \frac{v_{d - 1}}{8 \pi} \frac{(d - 1) (d - 2)}{(d - 3)} \Bigg) \frac{2 \Lambda}{\eta (d - 1)} r_v^2, 
\end{eqnarray}
and $M_{0, \text{trig}}$ and $M_{0, \text{exp}}$ are given by
\begin{eqnarray}
    M_{0, \text{trig(exp)}} &=& \frac{v_{(d - 1)}}{16 \pi} \frac{(d - 1)}{(d - 3)} (1 + w) \ell \kappa \rho_{0, \text{trig(exp)}} (r_b a_0)^{2/\ell}.
\end{eqnarray}

Since it is possible for a temporary trapped surface to occur in the trigonometric case as discussed in the previous subsection, it is instructive to check whether an apparent horizon might form in the exterior spacetime when the temporary trapped surface emerges. Using the exterior metric in Eq.~\eqref{eq:VaidyaMetric2}, the outgoing null geodesics can be obtained from
\begin{equation}
    0 = -f_\text{trig}(r_v) dV^2 - 2 dr_v dV,
\end{equation}
which then implies
\begin{equation}
    \frac{dr_v}{dV} = -\frac{f_\text{trig}(r_v)}{2}.
\end{equation}
The event horizon is formed when $\frac{dr_v}{dV} = 0$, which implies $f_\text{trig}(r_v) = 0$. Since $r_v = r_b a(t)$ [see Eq.~\eqref{eq:FirstFormMatch}], we can analyze the time the horizon is formed by studying the function $f_\text{trig}(r_v(t))$ in terms of $t$. Plotting the function $f_\text{trig}$ in Fig.~\ref{fig-ftrig} using the same parameters used in Fig.~\ref{fig-Atrig}, we find that the apparent horizon in the exterior spacetime is formed when the final state of the gravitational collapse is a black hole, but it is not formed when the final state is a naked singularity. Therefore, when the gravitational collapse to naked singularity takes place and the temporary trapping occurs, it is not the apparent horizon of the exterior spacetime that prevents the outgoing light rays from reaching distant observers, but rather the intricate collapse dynamics unfolding within the interior spacetime.

\section{Conclusions}

\label{sec-conclusions}

We have investigated possible final states of a spherically symmetric homogeneous perfect fluid undergoing a gravitational collapse in the framework of Rastall gravity. Following Ref.~\cite{Ziaie-et-al-gravitational-2019}, but now working in higher dimensions and including also the cosmological constant term, we study the structure of the singularity that is formed at the end of the collapse by mapping possible outcomes of the collapse in the space of the Rastall parameter $\gamma$ and the barotropic index $w$. The key to analysis here is a function $\ell$, which is a function of $\gamma$, $w$, and the spacetime dimension $d$ [see Eq.~\eqref{eq:ell}]. If $\ell > 1$, then the collapse will lead to a naked singularity formation, because in this case there is no apparent horizon which can cover the singularity. If $\ell < 1$, then the final state of the collapse is a black hole, because an apparent horizon will form before the end of the collapse such that the singularity is covered from distant observers. If $\ell < 0$, then there is no possible collapse which can lead to a singularity formation. Mapping these possible outcomes in Fig.~\ref{fig-parameter-space}, we have demonstrated that the naked singularity formation is possible in both the upper and lower branches of the parameter space $(w, \gamma)$ for any finite spacetime dimension $d$. However, as the dimension $d$ is increased, the naked singularity regions are shrinking before vanishing in the limit $d \to \infty$. Therefore, our results for any finite spacetime dimension $d$ provide counterexamples to the cosmic censorship conjecture.

The map depicted in Fig.~\ref{fig-parameter-space} of possible final states of the gravitational collapse in the parameter space $(w, \gamma)$ is the same for both cases of vanishing and nonvanishing cosmological constants, since the function $\ell$ does not depend on the cosmological constant $\Lambda$. However, when $\Lambda \neq 0$, there are two distinct solutions for the collapse evolution: trigonometric and exponential solutions. If $\Lambda$ and $\eta$, which is a function of $\gamma$ and $d$ [see Eq.~\eqref{eq:eta}], have the same sign (different signs), then the collapse evolution is in trigonometric (exponential) form. In the case of trigonometric (exponential) solution, the effective energy density $\rho_\text{eff}$ of the perfect fluid star is lower (higher) than the one in the case of $\Lambda = 0$ [see Eqs.~\eqref{eq:rhoeffCosmoTrig} and \eqref{eq:rhoeffCosmoExp}]. Since higher $\rho_\text{eff}$ means a stronger gravitational attraction in the collapse process, it implies that the gravitational collapse in the case of trigonometric (exponential) solution is slower (faster) than the one in the case of $\Lambda = 0$ (see Fig.~\ref{fig-DCollapse-d}).

Besides the difference in the collapse duration, there are at least two other reasons why trigonometric and exponential solutions are physically distinct. First, for the collapse to be possible in the trigonometric solution in both the upper and lower branches of the parameter space and in both the black hole and naked singularity regions in each branch, the magnitude of $\Lambda$ cannot exceed some cutoff value $| \Lambda_\text{cut} |$. Otherwise, the effective energy density $\rho_\text{eff}$ will be negative, which is unphysical. This is due to the fact that increasing $| \Lambda |$ will decrease $\rho_\text{eff}$ in the trigonometric case [see Eq.~\eqref{eq:rhoeffCosmoTrig}]. Therefore, for a point in the upper (lower) branch of the parameter space, where $\eta$ is positive (negative), either in the black hole or the naked singularity region, the collapse is not possible to occur when $\Lambda > \Lambda_\text{cut} > 0$ ($\Lambda < \Lambda_\text{cut} < 0$). In contrast, there is no cutoff value for $\Lambda$ in the exponential solution, since in this case increasing $| \Lambda |$ will only increase $\rho_\text{eff}$ [see Eq.~\eqref{eq:rhoeffCosmoExp}].

Second, we have demonstrated that it is possible for a temporary trapped surface to emerge in the case of trigonometric solution in the naked singularity region only. This occurs when the magnitude of $\Lambda$ exceeds some critical value $| \Lambda_\text{crit} |$ but less than the cutoff value $| \Lambda_\text{cut} |$. Since $| \Lambda_\text{cut} | > | \Lambda_\text{crit} |$ (see Fig.~\ref{fig-DCollapse-Lambda}), the emergence of the temporary trapped surface is not ruled out by the positivity condition of $\rho_\text{eff}$. Hence, for a point in the naked singularity region in the upper (lower) branch of the parameter space, where $\eta$ is positive (negative), a temporary trapped surface can emerge when $0 < \Lambda_\text{crit} < \Lambda < \Lambda_\text{cut}$ ($\Lambda_\text{cut} < \Lambda < \Lambda_\text{crit} < 0$). When this occurs, the collapsing object will not be visible to distant observers because all outgoing radial light rays are trapped. Hence, if they observe the collapse process in a timescale shorter than the collapse duration, they may conclude that a black hole is formed, although the collapse will eventually lead to a naked singularity formation. We also find that there is no event horizon which forms in the exterior spacetime when the temporary trapping occurs (see Fig.~\ref{fig-ftrig}). Therefore, light rays are prevented from reaching distant observers not due to the presence of an apparent horizon in the exterior spacetime, but rather as a consequence of the complex collapse dynamics occurring within the interior spacetime.

\section*{Acknowledgments}

We thank the late Muhammad Iqbal for his contribution in the early calculations of this work. This research was supported by the ITB Research Grant.

\bibliography{references}

\end{document}